\documentclass{article}
\usepackage{spconf,amsmath,graphicx}
\usepackage{amsfonts}
\usepackage{hyperref}

\title{IN SILICO PREDICTION OF CELL TRACTION FORCES}
\name
  {Nicolas Pielawski$^{\star}$, Jianjiang Hu$^{\diamond}$, Staffan Str\"{o}mblad$^{\diamond}$,Carolina W\"ahlby$^{\star}$\thanks{Thanks to the Swedish Foundation for Strategic Research for funding (grant SB16-0046).}}
\address{$^{\star}$Uppsala University, Dept. of Information Technology, L\"{a}gerhyddsv\"{a}gen 2, 752 37 Uppsala\\
    $^{\diamond}$Karolinska Institutet, Department of Biosciences and Nutrition, H\"{a}lsov\"{a}gen 7C, 141 57 Huddinge}
\begin{document}
%
\maketitle
\begin{abstract}
Traction Force Microscopy (TFM) is a technique used to determine the tensions that a biological cell conveys to the underlying surface. Typically, TFM requires culturing cells on gels with fluorescent beads, followed by bead displacement calculations. We present a new method allowing to predict those forces from a regular fluorescent image of the cell. Using Deep Learning, we trained a Bayesian Neural Network adapted for pixel regression of the forces and show that it generalises on different cells of the same strain. The predicted forces are computed along with an approximated uncertainty, which shows whether the prediction is trustworthy or not. Using the proposed method could help estimating forces when calculating non-trivial bead displacements and can also free one of the fluorescent channels of the microscope. Code is available at \url{https://github.com/wahlby-lab/InSilicoTFM}.
\end{abstract}
\section{Introduction}
\label{sec:intro}

In 1999, Dembo et al. developed a new technique named Traction Force Microscopy and made possible the visualisation of cellular forces by placing cells on a soft gel containing randomly placed fluorescent beads\cite{dembo1999stresses}. Using this technique, they could retrieve the displacement of the beads and generate a vector field representing the local forces, with their magnitude and direction. Traction Forces are often used in studies of cell migration patterns; it has been hypothesised that cancer cells exhibiting high traction forces are more invasive than cells with lower activity \cite{koch20123d}.

In 2010, Lemmon et al.\cite{lemmon2010predictive} -- on the basis that the shape and size of a cell correlates with the amount of forces it exerts on the substrate -- described a method allowing the prediction of traction forces based solely on the cell geometry. 

In recent years, neural networks gained popularity in research in biology mainly due to their successful application to various problems as well as the availability of computing capacity to train them. In 2015, Ronneberger et al.\cite{ronneberger2015u} proposed a new type of neural network architecture: the U-Net. It contains a down-sampling path working on feature maps of decreasing resolutions and followed by an up-sampling path that constructs the output image. Both paths are connected with skip-connections that allow information to flow through the network. In 2017, J\'{e}gou et al. built upon the U-Net architecture to create the Tiramisu neural network\cite{jegou2017one}. Adding various improvements, such as the dense blocks from the DenseNet neural network by Huang et al.\cite{huang2017densely}, they could improve the performances of the U-Net architecture while substantially reducing the number of trainable weights.

The insight from the method by Lemmon et al.\cite{lemmon2010predictive}, combined with a neural network able to use the structural information of a cell at different scales such as the Tiramisu network, offers an opportunity to translate a fluorescent image of a cell into an image estimating traction forces. To the best of our knowledge, this paper describes the first attempt at using deep neural networks to predict cell traction forces.
 
This paper describe the process that leads to the creation of the data, the construction of the deep neural network to predict the cellular forces and the measurement of the uncertainty around the prediction of the neural network.

\begin{figure*}[htb]
    \center
    \includegraphics[width=\textwidth]{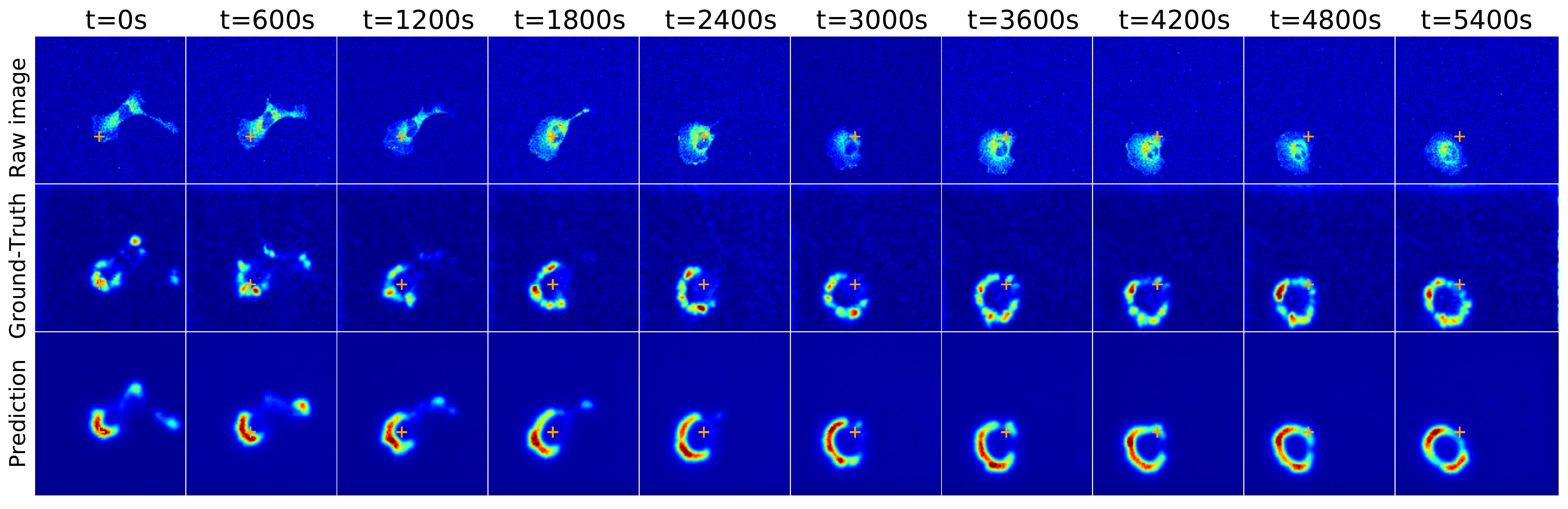}
    \caption{Example raw images fed to the neural network (top row), the ground truth (middle row), and the predicted forces (bottom row) of the test cell over a time series. The orange crosses represent the pixel chosen for generating figure \ref{fig:pixelpred}. The full sequence -- the ground-truth force overlapped with the prediction of the test cell -- is available at \url{https://youtu.be/QhzNmrA42T4}.}
    \label{fig:frames}
\end{figure*}

\section{Material and Methods}
\label{sec:material_n_methods}
\subsection{Image data}
\label{ssec:image_data}
 Immortalised human fibrosarcoma cell line HT1080 stably expressing a FRET based RhoA biosensor\cite{Fritz2013FRET} was used in this study. The cells were seeded on collagen type I coated with red fluorescent beads (Invitrogen, F8801) containing polyacrylamide gel (6.9 kPa) three hours before the imaging started. An environmental chamber equipped with a Nikon A1 confocal microscope with 60x oil objective (NA 1.4) was used to image single-cell migration and displacements of the fluorescent beads at a resolution of 200 nm/pixel and a time interval of 30 seconds for 1.5 hours. 457 nm and 561 nm lasers were used to excite the FRET biosensor and red fluorescent beads respectively, while 525/50 and 595/50 emission filters were used to collect the signals. After the time-lapse imaging, cells were trypsinized and single snapshots of fluorescent beads were collected to get the positions of the beads at the released state. MATLAB R2014b with the traction force microscopy package from Danuser Lab. \cite{Han2015TFM} was used to calculate the traction force based on the bead displacements.

Two datasets were generated in this fashion at two different occasions and are available on Zenodo\cite{jianjianghu_2019}. The first dataset (12 cells) was used for training and the second one (11 cells, and 1 test cell) for validation purposes. The last cell from the validation dataset was taken out to generate the figures and will be called test cell.

\begin{figure*}[htb]
    \includegraphics[width=\textwidth]{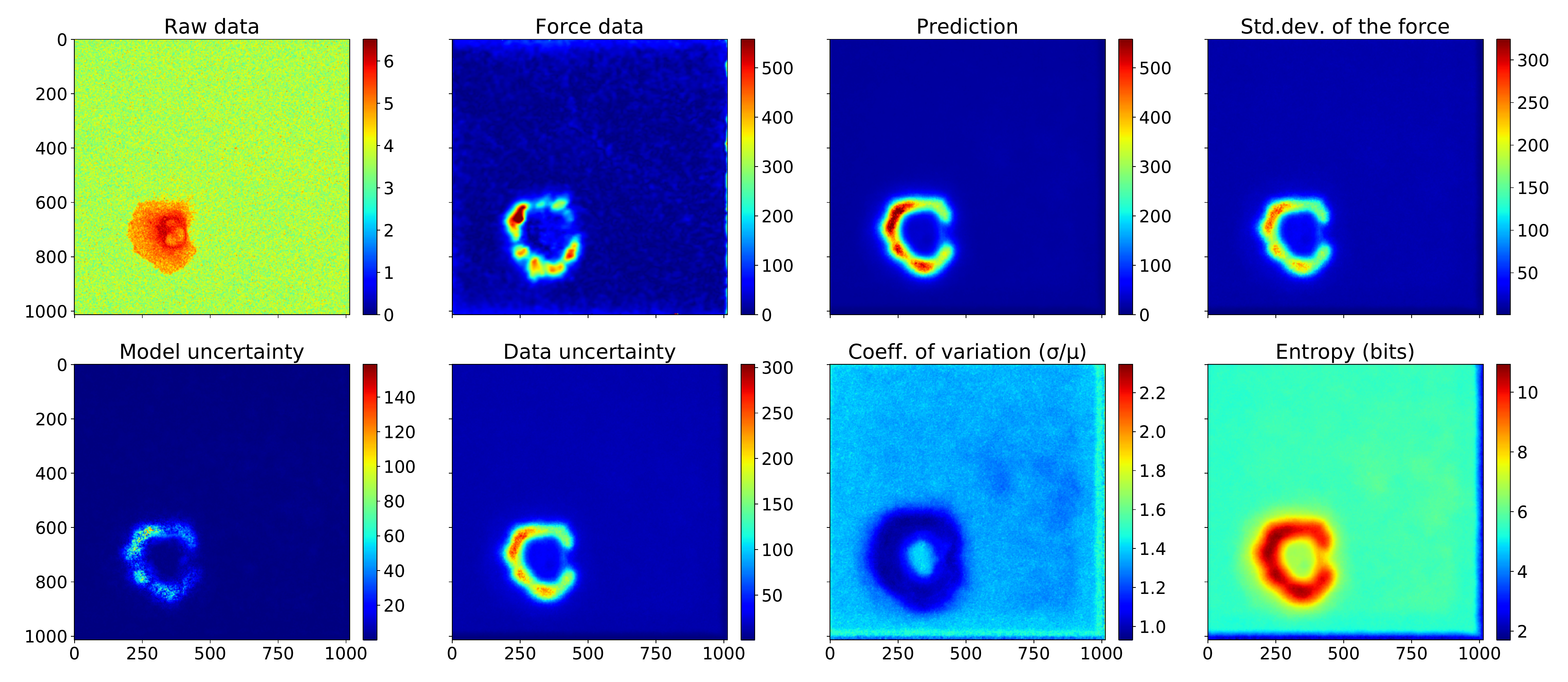}
    \caption{The test cell at time 4320s. The raw data represents the data that was fed to the neural network. The remaining images have been generated from the output of the neural network. The full time sequence is available at: \url{https://youtu.be/U9-Tn9ojXAU}.}
    \label{fig:prediction}
\end{figure*}

\subsection{Deep Learning Model}
\label{ssec:dlmodel}

We modified a Tiramisu segmentation neural network\cite{jegou2017one} -- A U-Net architecture made of Dense blocks from the DenseNet architecture -- in order to predict forces and their uncertainties.

The last layer has been replaced by two fully connected layers in order to predict the mean and the variance (aleatoric uncertainty) of the estimated forces. The mean forces have no activation function (linear activation) while the variance has a softplus activation squared: $\ln(1+\exp(x))^2$. The weights of the neural networks are initialised with a Glorot normal distribution except for the variance layer, which is initialised with zeros for the kernels, and $\ln(e-1)$ for the biases, so that the variance is 1 at the beginning of the training.

The drop-out layers have been modified so that they are active during training as well as during inference, in order to predict the epistemic uncertainty. The rate of drop-out is 20\% throughout.

The architecture consists of six dense blocks, with five layers per block and a growth rate of 16. The first initial filter has a size of 3x3 and a depth of 48. In the expansive path of the neural network, at each level, the feature maps are up-sampled by a factor of two without interpolation followed by 2-D convolutions with a depth of 128 filters. The Adam optimiser is used with L-2 weight decay of $10^{-4}$ and gradient clipping (maximum L-2 norm of $1.0$).

The model was trained on four Titan Xp graphic cards over 200 epochs, with 50 steps per epoch and a batch size of 8, so that each GPU deals with two images at a time.

\subsubsection{Loss functions}
\label{sssec:loss}
The Kullback-Leibler (KL) divergence measures the relative entropy between two probability distributions. Training a neural network by Maximum Likelihood Estimation (MLE) is analogous to minimising the KL divergence. Because the log of the data is normally distributed, we used the KL divergence between two log-Normal distributions that yields the following loss function\cite{gil2013renyi}:
\begin{multline}
    \mathcal{L}_\textrm{MSE}(\boldsymbol{\mu}, \boldsymbol{\hat{\mu}}, \boldsymbol{\hat{\sigma}^{2}}) = D_{KL}(\ln \mathcal{N}(\boldsymbol{\mu}, 0) || \ln \mathcal{N}(\boldsymbol{\hat{\mu}}, \boldsymbol{I_{n}\hat{\sigma}^{2}})) \\
    \propto \frac{1}{n} \sum_{i=0}^{n} \left( \frac{1}{2\hat{\sigma}_i^2} ||\mu_i - \hat{\mu}_i||^2 + \frac{1}{2}\ln{\hat{\sigma}_i^2} \right)
\end{multline}
with $n$ the number of data points, $\boldsymbol{\mu}$ the ground truth, $\boldsymbol{\hat{\mu}}$ the prediction, $\boldsymbol{\hat{\sigma}^2}$ the predicted uncertainty, and assuming we have no uncertainty on the ground-truth.

This loss function is the mean squared error loss that would be derived from the KL divergence between two normal distributions. Due to the properties of the log-normal distribution, the ground-truth data needs to be log-transformed and the trained neural network will yield log-forces.

\subsection{Image augmentation}
Construction of the forces based on the bead displacement creates high intensity artefacts close to the borders. To cope with this unwanted behaviour all images of forces were masked with a 2-dimensional 10\% cosine-tapered (Tukey) window\cite{harris1978use}.

The construction of the batches follows a pipeline with three distinct steps. First, because the images are composed of mainly background, we extracted the cells from the images by thresholding them. A morphological closing and opening were applied sequentially with a box kernel of size 5x5. A bounding box was then fitted around the biggest blob available. In the second step, the images were flipped horizontally then vertically with a probability of 50\%. Resulting images were randomly rotated with a bi-cubic interpolation and randomly cropped to create uniform batches of size 256x256. Salt noise was added over 1\% of the pixels of 50\% of the input images, with a random intensity sampled from a uniform  distribution ($a=0, b=2000$). This increased the robustness of the neural network towards sparse noise of potential high intensity. The final step consisted of log-transforming the resulting images with the clipped log function: $\ln(\max(1, x))$.

\subsection{Measure of the uncertainty of the predictions}
Measuring uncertainty is an important factor of this study, as it adds another dimension to the understanding of the output of the neural network. Indeed, allowing users to know whether a prediction can be trusted, and to which extent, can be useful for further research. For instance, it becomes possible to select images that reach only a specific certainty, or even perform statistical testing.

\begin{figure}[htb]
    \includegraphics[width=\linewidth]{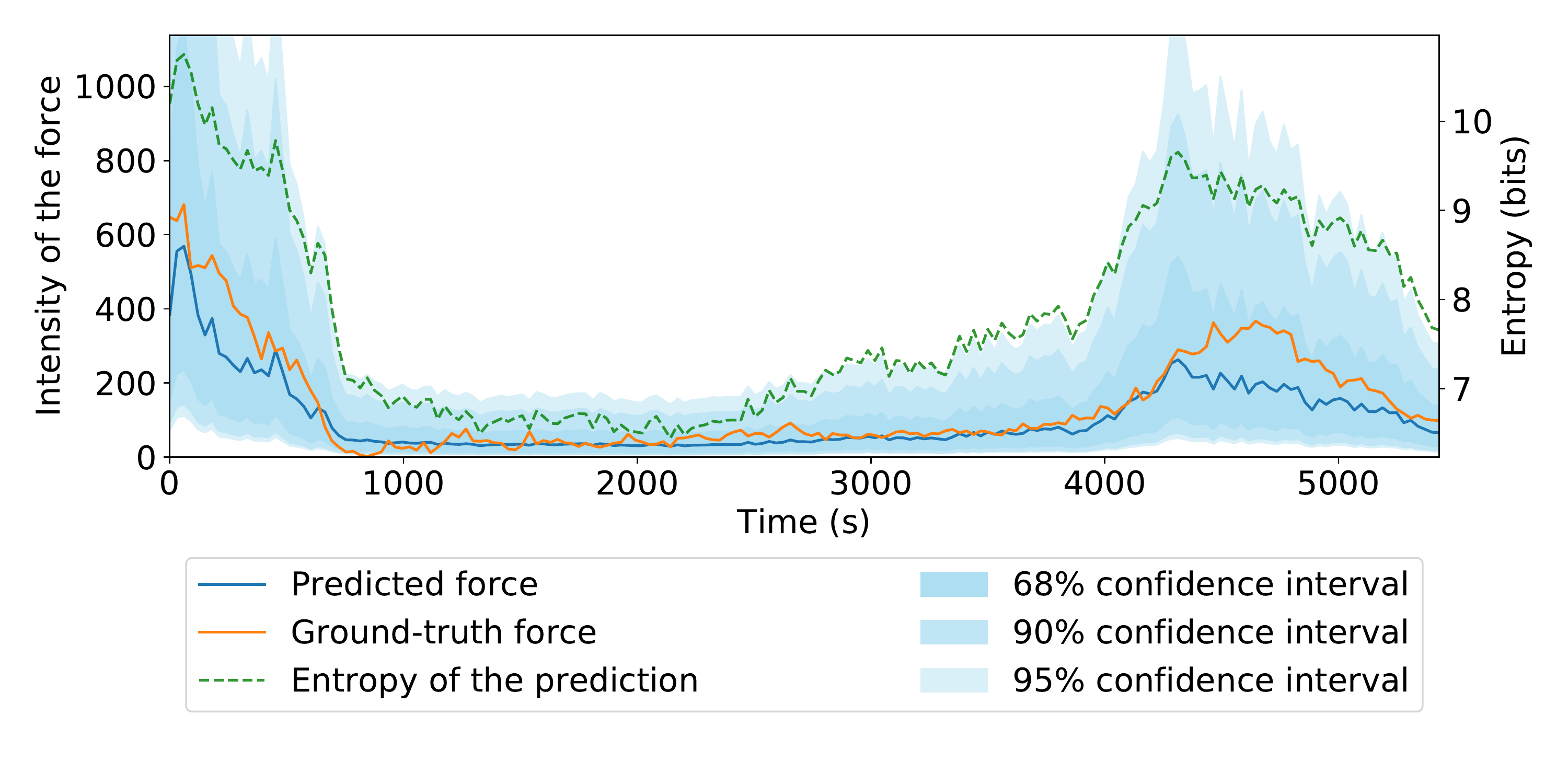}
    \caption{Ground-truth and predicted force of an arbitrary pixel (displayed as an orange cross in figure \ref{fig:frames}). Blue regions represent prediction confidence intervals, and entropy is displayed as a dashed line.}
    \label{fig:pixelpred}
\end{figure}

\subsubsection{Aleatoric and Epistemic uncertainties}
\label{sssec:uncertainty}
The aleatoric and epistemic are two different ways to communicate about uncertainties. The former can be quantified in closed form and provides an uncertainty related to the lack of information in the input data, whereas the latter needs to be approximated and yields information about the lack of agreement within the model. The approximation of the epistemic uncertainty was achieved using the Monte Carlo drop-out method as described by Kendall et al. in \cite{kendall2017uncertainties}, with some modifications to accommodate for the log-normal distribution.

Given that a neural network $t$ has the ability to formulate a prediction
$\mu_t$ and aleatoric uncertainty $\sigma_t^2$, the force can be computed as follows:
\begin{equation}
    \mathbb{E}[\hat{y_i}|x_i^*,\textbf{X}] \approx \frac{1}{T} \sum_{t=1}^{T} \exp(\mu_t + \sigma_t^2 / 2)
\end{equation}
for a given pixel $x_i^*$ and input image $\textbf{X}$, and $T$ sampled neural networks where
the weights are sampled from a drop-out distribution.

Accordingly, the full prediction variance is derived as:
\begin{multline}
    \mathbb{V}[\hat{y_i}|x_i^*,\textbf{X}] \approx \frac{1}{T} \sum_{t=1}^{T}
    (\exp(2\mu_t + \sigma_t^2)(\exp(\sigma_t^2) - 1) + \\
    \frac{1}{T} \sum_{t=1}^{T} \exp(\mu_t + \sigma_t^2 / 2)^2 - \mathbb{E}[\hat{y_i}|x_i^*,\textbf{X}]^{2}
\label{eq:var}
\end{multline}

This variance is a sum consisting of both the aleatoric and the epistemic uncertainties, respectively.

\subsubsection{Coefficient of Variation}
The coefficient of variation is derived by dividing the standard deviation by the mean:
\begin{equation}
    CV[\hat{y_i}|x_i^*,\textbf{X}] \approx \frac{1}{T} \sum_{t=1}^{T}\sqrt{\exp{\sigma_t^2} - 1}
\end{equation}
and gives information about the amount of uncertainty in relation to the intensity of the force, even though the formula does not take the parameter $\mu$ into account.

\subsubsection{Differential Entropy}
The entropy of a log-normal distribution is defined in Kvalseth
\cite{kvalseth1982some}, and can be approximated in the following way:
\begin{multline}
    \mathbb{H}[\hat{y_i}|x_i^*,\textbf{X}] \approx \frac{1}{T} \sum_{t=1}^{T} \log_2(\sigma_t\exp(\mu_t+\frac{1}{2}) /\sqrt{2 \pi})
\end{multline}
and yields the predictive entropy of the neural network. This method, used by Nair et al. ``is a measure of how much information is in the model predictive density function at each [pixel] $i$''\cite{nair2019exploring}. This measure reveals the relative number of bits missing from each individual log-normally distributed pixel.

\section{Results}
\label{sec:results}

\begin{figure}[htb]
    \includegraphics[width=\linewidth]{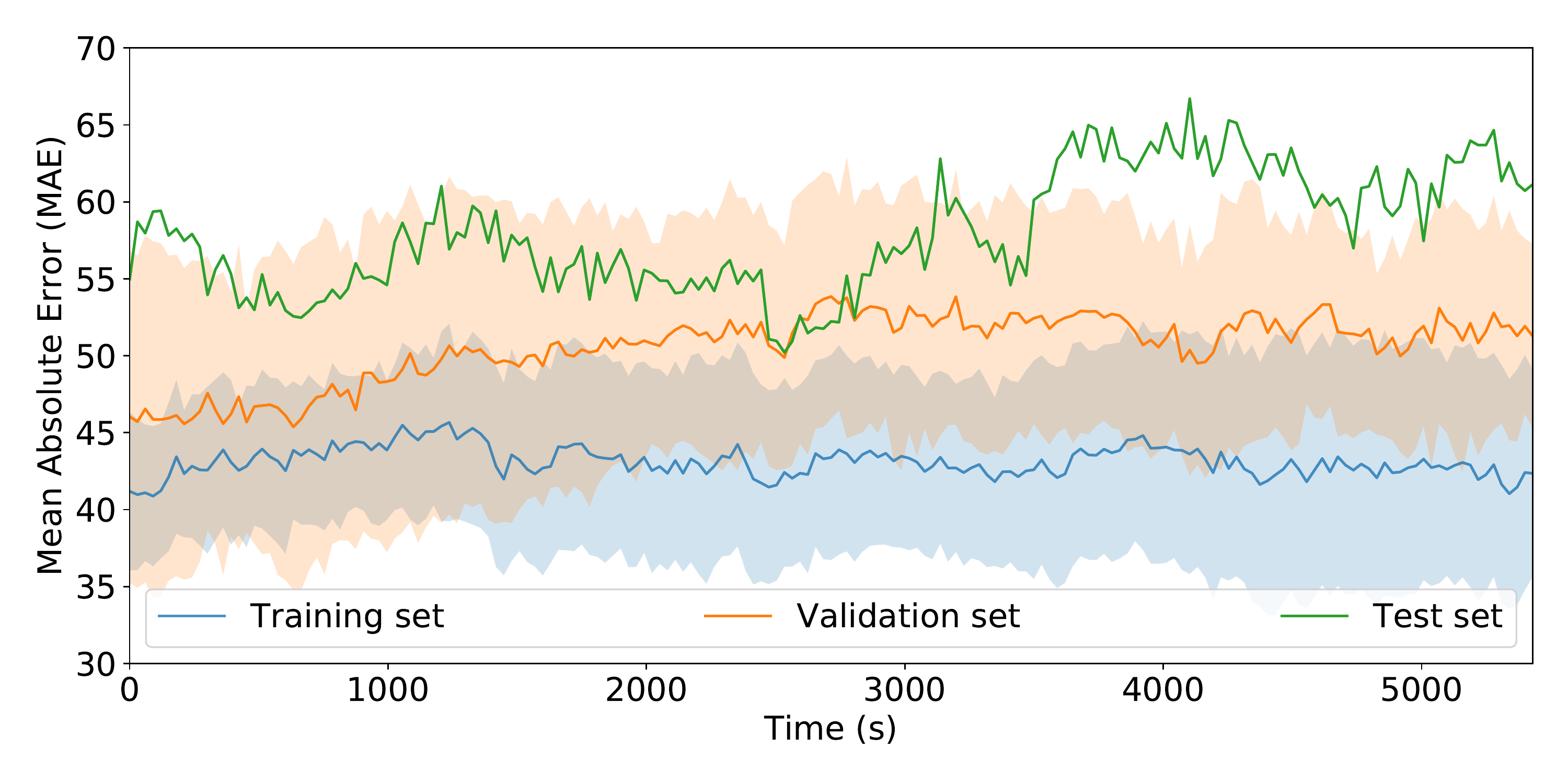}
    \caption{Mean Absolute Error (MAE) of the training, validation and test sets for each individual frame. The standard
    deviation represents the spread around the mean for the training set (12
    cells) and the validation set (11 cells).}
    \label{fig:maeovertime}
\end{figure}

The neural network was evaluated on the validation set; the test set was used for illustration purposes only, the validation set was not used for hyper-parameter optimisation in order to avoid a possible human bias.

Figure \ref{fig:frames} shows the predicted forces of the test cell over time. A small orange cross represents a pixel chosen arbitrarily to generate Figure \ref{fig:pixelpred}.

Figure \ref{fig:pixelpred} represents the prediction of the forces compared to the ground-truth. The log-normal distribution percent-point (quantile) function was used to generate confidence intervals.

Figure \ref{fig:maeovertime} displays the Mean-Absolute Error over the time sequence of the cells (181 frames) for each of the sets. The sets were not augmented, the 10\% Tukey mask was still applied to the forces. The mean MAE of the training set reached $43.12 \pm 5.49$ (1 standard deviation), $50.57 \pm 7.13$ for the validation set and $58.05$ on the test cell. The figure shows some signs of over-fitting as the training set MAE outperforms the testing and validations sets. Globally, the error remains stable over time, even though the i.i.d. assumption of the data was not respected.

\section{Discussion and Conclusion}

We presented a novel method: using Deep Learning to predict cellular traction forces, even if the information directly related to the forces is not available. Indeed, despite the neural network not having access to the beads or fluorescent channels linked to proteins correlated with the force, it successfully made use of the cell geometry to accurately infer cell forces.

Adding channels representing fluorescent proteins related to cellular traction forces -- i.e. actin or integrin -- could bring dramatic improvements to the accuracy of the Deep Learning model. More, relying on the intensity of a fluorescent protein could increase the generalisation and stability when varying the softness of the gel, or using a glass medium. In addition, our method has so far been applied to one cell line and generalisation to other cell lines will require further testing and development.

\bibliographystyle{IEEEbib}
\bibliography{refs}

\end{document}